\documentclass[11pt,a4paper]{article}
\pdfoutput=1
\usepackage{jcappub}

\usepackage{amssymb} 
\usepackage{mathrsfs} 
\usepackage{multirow} 
\usepackage{subfig} 
\usepackage{bm} 
\usepackage{graphicx}
\usepackage{amsmath,color}
\usepackage{amsfonts}

\newcommand{\Mnu}{$\Sigma_i m_{\nu_i}~$}

\newcommand{\mincir}{\raise
  -2.truept\hbox{\rlap{\hbox{$\sim$}}\raise5.truept \hbox{$<$}\ }}
\newcommand{\magcir}{\raise
  -2.truept\hbox{\rlap{\hbox{$\sim$}}\raise5.truept \hbox{$>$}\ }}
\newcommand{\siml}{\raise
  -2.truept\hbox{\rlap{\hbox{$\sim$}}\raise5.truept \hbox{$<$}\ }}
\newcommand{\simg}{\raise
  -2.truept\hbox{\rlap{\hbox{$\sim$}}\raise5.truept \hbox{$>$}\ }}

\begin{document}

\title{Cosmology with massive neutrinos III: the halo mass function and  
an application to galaxy clusters}

\author[a,b]{Matteo Costanzi,}
\author[c]{Francisco Villaescusa-Navarro,}
\author[c,b]{Matteo Viel,}
\author[d]{Jun-Qing Xia,}
\author[a,b,c]{Stefano Borgani,}
\author[e]{Emanuele Castorina,}
\author[f,g]{Emiliano Sefusatti}

\affiliation[a]{Universit\`{a} di Trieste, Dipartimento di Fisica,\\ via Valerio, 2, 34127 Trieste, Italy}
\affiliation[b]{INFN-National Institute for Nuclear Physics,\\ via Valerio 2, 34127 Trieste, Italy}
\affiliation[c]{INAF-Osservatorio Astronomico di Trieste,\\ via Tiepolo 11, 34133 Trieste, Italy}
\affiliation[d]{Key Laboratory of Particle Astrophysics, Institute of High Energy Physics, Chinese Academy of Science,\\ P.O. Box 918-3, Beijing 100049, People’s Republic of China}
\affiliation[e]{SISSA - International School For Advanced Studies, Via Bonomea, 265 34136 Trieste, Italy}
\affiliation[f]{The Abdus Salam International Center for Theoretical Physics,\\Strada Costiera 11, 34151, Trieste, Italy}
\affiliation[g]{INAF, Osservatorio Astronomico di Brera,\\ Via Bianchi 46, I-23807 Merate (LC)  Italy}

\emailAdd{costanzi@oats.inaf.it}
\emailAdd{villaescusa@oats.inaf.it}
\emailAdd{viel@oats.inaf.it}
\emailAdd{xiajq@ihep.ac.cn}
\emailAdd{borgani@oats.inaf.it}
\emailAdd{castori@sissa.it} 
\emailAdd{emiliano.sefusatti@brera.inaf.it}

\abstract{We use a suite of N-body simulations that incorporate
  massive neutrinos as an extra-set of particles to investigate 
  their effect on the halo mass function.
  We show that for cosmologies with massive neutrinos  
  the mass function of dark matter haloes selected
  using the spherical overdensity (SO) criterion 
  is well reproduced by the fitting formula of
  Tinker et al. (2008) once the cold dark matter power spectrum is considered
  instead of the total matter power, as it is usually done.  The
  differences between the two implementations, i.e. using $P_{\rm
    cdm}(k)$ instead of $P_{\rm m}(k)$, are more pronounced for large values
  of the neutrino masses and in the high end of the halo mass function:
   in particular, the number of massive
  haloes is higher when $P_{\rm cdm}(k)$ is considered rather than  $P_{\rm m}(k)$. 
  As a quantitative application of our findings we consider a \textit{Planck}-like 
  SZ-clusters survey and show that the differences in predicted number counts can be as large as $30\%$ for $\sum m_\nu = 0.4$ eV. Finally, we use the \textit{Planck}-SZ clusters sample, with an approximate likelihood calculation,
  to derive \textit{Planck}-like constraints on cosmological parameters.
  We find that, in a massive neutrino cosmology, our correction to the halo mass function
  produces a shift in the $\sigma_8(\Omega_{\rm m}/0.27)^\gamma$ relation
  which can be quantified as $\Delta \gamma \sim 0.05$ and 
  $\Delta \gamma \sim 0.14$ assuming one ($N_\nu=1$) or three ($N_\nu=3$) degenerate massive neutrino, respectively.
  The shift results in a lower mean value of $\sigma_8$ with 
  $\Delta \sigma_8 = 0.01$ for $N_\nu=1$ and $\Delta \sigma_8 = 0.02$ for $N_\nu=3$, respectively. 
  Such difference, in a cosmology with massive neutrinos, would increase the tension between cluster abundance
  and \textit{Planck} CMB measurements.}


\maketitle

\section{Introduction}\label{sec_int}

Neutrinos are spin one-half leptons carrying no electric charge. 
Within the particle standard model they are described as elementary massless particles. 
Measurements of the $Z$ boson lifetime have pointed out that the 
number of active neutrinos is 3 ($N_\nu^{\rm active} = 2.9840 \pm 0.0082$, \cite{2005hep.ex....9008T}). 
On the other hand, the neutrino oscillation phenomenon
indicates that at least two of the three neutrino families have to be massive. 
Unfortunately, measurements involving neutrino flavour changing only 
provide us with information about the mass square differences between the different 
mass eigenstates, i.e. they can not be used to determine the absolute neutrino 
mass scale. Recent experiments using solar, atmospheric and reactor neutrinos quantified these differences as: $\Delta m_{12}^2=7.5 \times 10^{-5}\, \text{eV}^2$ and
$|\Delta m_{23}^2|=2.3 \times 10^{-3}\, \text{eV}^2$ (see e.g. \cite{Fogli_2012, 2012PhRvD..86g3012F}), where $m_1$, $m_2$ and $m_3$ are the masses of the different neutrino mass eigenstates.
Since we are not capable to measure the sign of $\Delta m_{23}^2$, two different mass ordering (hierarchies) are possible: a normal hierarchy ($m_2 < m_3$) and an inverted hierarchy ($m_2 > m_3$).
Therefore, the sum of the neutrino masses is constrained from below as \Mnu $>$ 0.056, 0.095 eV  depending on whether neutrinos follow the normal or inverted hierarchy, respectively. 
Knowing the absolute neutrinos mass scale is of great importance, since it is related 
to physics beyond the particle standard model. For this reason, a huge effort from both
the theoretical and the experimental side is currently on-going with the purpose to weight neutrinos.

From the cosmological point of view, the Big Bang theory predicts the existence of a cosmic
neutrino background 
(see, e.g. \cite{2002PhR...370..333D, 2006PhR...429..307L} for a review). In the very early Universe, cosmic neutrinos contributed to the total radiation energy density, affecting the nucleosynthesis process and therefore the primordial abundance of light elements. At the linear order, massive neutrinos impact cosmology in different ways, depending on which parameters are fixed: they shift the matter-radiation equality time, at fixed $\Omega_{\rm m}$, and they slow down the growth of matter perturbations during the matter and Dark Energy dominated era.
 The combination of the above two effects produces a suppression in the amplitude of 
the matter power spectrum on small scales (see for instance \cite{2006PhR...429..307L}).

The imprints left by massive neutrinos on the CMB and on the Large Scale Structucture (LSS)
of the Universe
have been used to set upper limits on their masses. Numerous recent works point towards
neutrino masses, $\Sigma_i m_{\nu_i}$, below 0.3 eV at $2\sigma$ \citep[e.g.][]{2012arXiv1211.3741Z, Xia_2012, 2012arXiv1210.2131R,
2012arXiv1202.0005J,Planck_cosm_parameters}, with the notable exceptions of
\citep{2006JCAP...10..014S} in which authors used Lyman$-\alpha$ data to set an upper
limit of $0.17\, \text{eV}$ and \cite{Riemer-Sorensen} who found \Mnu $< 0.18$ eV ($95\%$) by 
combining data from BAO, CMB and the WiggleZ galaxy power spectrum\footnote{Private communication}.

Great attention has been recently drawn to the tension between the \textit{Planck} 
measurements of the primary CMB temperature anisotropies~\cite{Planck_cosm_parameters}
and measurements of the current expansion rate $H_0$~\cite{Riess2011}, the galaxy shear power 
spectrum~\cite{CFHTLenS2013}
and galaxy cluster counts~\cite{2009ApJ...692.1060V,2010ApJ...708..645R,PlanckSZ2013}.
Besides unresolved systematic effects, it has been suggested by many authors~
\cite{PlanckSZ2013,Wyman2013,Battye2013,Hamann2013}
that the discrepancy can be alleviated by extending
the standard $\Lambda$CDM model to massive neutrinos, either active or sterile.
A common finding of those works is that a neutrino mass of $0.3-0.4 ~{\rm eV}$ 
provides a better fit to the combination of CMB data and low redshift Universe 
measurements than the vanilla $\Lambda$CDM model.

Among the different probes of the LSS, galaxy clusters have played a significant role in the definition of the ``concordance'' $\Lambda$CDM model~\citep[e.g.][]{2011ARA&A..49..409A,2012ARA&A..50..353K}, and many ongoing (Planck, SPT, DES), upcoming and future (eROSITA, LSST, Euclid) surveys will aim to use their abundances and spatial distribution to strongly constrain cosmological parameters. In order to fully exploit for cosmology the ever growing number of clusters detected, it is mandatory to have a reliable theoretical predictions for the cluster abundance (the halo mass function, HMF)~\cite{Reed2013, Crocce_2010, Watson}, together with an accurate calibration of the observable-mass relation. As for the former, since the pioneering work of Press \& Schechter~\cite{PS1974} many forms for the HMF have been proposed in literature
~\citep[e.g.][]{Sheth-Tormen, Jenkins, Reed, Warren, Tinker2008, Crocce_2010}, often
 calibrated against large suites of cosmological simulations. 
Despite the great improvement of the numerical results over the past decade many sources of 
systematic error still affect the HMF, including
finite simulation volume, mass and force resolution, baryonic physics and massive neutrino effects. Here we focus on the consequences of non-vanishing neutrino masses.
 
The effects of neutrino masses on the halo mass function has already been studied in different works \cite{Brandbyge_halos, Marulli_2011, Ichiki-Takada, Paco_2013a}. 
In the work of \cite{Brandbyge_halos}, the authors measured the halo mass function 
from N-body simulations incorporating massive neutrinos using a hybrid scheme to simulate neutrino particles. They showed that the halo mass function in models with massive neutrinos can be well reproduced by the Sheth and Tormen (ST) \cite{Sheth-Tormen} mass function by using 
$\rho_{\rm cdm}=\rho_{\rm m}-\rho_\nu$, instead of $\rho_{\rm m}$, when establishing the relation between the halo mass and the top-hat window function radius ($M=4\pi\rho R^3/3$, see section \ref{sec:massf} for details). Those results were later independently verified in \cite{Marulli_2011, Paco_2013a} using a different set of N-body simulations. More recently, the authors of \cite{Ichiki-Takada} investigated the gravitational collapse of a spherical region in a massive neutrino cosmology, showing that neutrinos play a negligible role in the process. This led to the conclusion that the cold dark matter power spectrum should be used to compute the r.m.s. of the matter perturbations, $\sigma(M)$, required predict the halo mass function. In \cite{Castorina_2013} this was tested againts N-body simulations, resulting in an excellent agreement. 
 
This paper is the last of a series of three papers. Paper I \cite{Paco_2013} introduces a large set of numerical simulations incorporating massive neutrinos as particles. It then studies the effect of neutrino masses on the spatial distribution of dark matter haloes, finding that halo bias, as typically defined w.r.t. the underlying {\em total} matter distribution, exhibits a scale-dependence on large scales for models with massive neutrinos. In addition, Paper I investigates as well massive neutrinos effects on the spatial distribution of galaxies by constructing mock galaxy catalogues using a simple halo occupation distribution (HOD) model.

In Paper II \cite{Castorina_2013} the universality of the HMF and of linear bias in massive neutrino cosmologies is discussed in terms of halo catalogues determined with the Friends-of-Friends algorithm on the simulations introduced in Paper I. It is shown that the proper variable to describe the HMF of a massive neutrino model is the variance of {\em cold} dark matter perturbations, rather than the total ones (i.e. including neutrinos) typically assumed in previous analyses \cite{Brandbyge_halos, Marulli_2011, Paco_2013a}. If the correct prescription is used then the HMF becomes nearly universal with respect to the neutrino mass. The paper discusses also similar results for the bias of haloes at large scales, which is found to be almost scale independent and universal when expressed in terms of CDM quantities alone.

In this paper we explore how the results of Paper II affect the determination of
cosmological parameters from galaxy clusters data. Here we study the HMF of dark matter
haloes identified using the Spherical Overdensity (SO) algorithm.
The reason to use SO haloes is that the mass proxy in X-ray
and SZ measurements is calibrated with spherically defined objects rather than
with the Friends-of-Friend (FoF) haloes considered in Paper II. We show that the abundance
of SO haloes is well reproduced by the Tinker fitting formula once the cold dark matter 
mean density and linear power spectrum are used, in agreement with Paper II and the work of
\cite{Ichiki-Takada}. Then, we show that
our findings have interesting implications  for cosmology using cluster number counts, 
especially due to the recently highlighted tension between cosmological parameter constraints
inferred from CMB temperature data and
the SZ clusters datasets \cite{PlanckSZ_catalogue}. As a case study, we choose the 
Planck SZ-selected sample of clusters~\cite{PlanckSZ_catalogue}, for which we perform a
Monte Carlo Markov Chain analysis in order to compare constraints obtained using 
different prescriptions for the halo mass function. We find that using the CDM linear matter
power spectrum when computing the r.m.s. of the smoothed linear density field, $\sigma(M,z)$ (i.e.
when using a better description for the HMF in massive neutrinos cosmologies) 
changes the degeneracy direction between the parameters
$\Omega_{\rm m}$ and $\sigma_8$ and decreases the $\sigma_8$ mean value.
These changes increase the tensions between cosmological parameters 
constraints from CMB data and from SZ cluster counts.

The paper is organized as follows. In Section~\ref{sec:sims} we
describe the numerical simulations we have used to calibrate the HMF of
dark matter haloes identified using the SO criterion.
The halo mass
functions for the different cosmological models and the procedure used
to compute them are shown in section \ref{sec:massf}. The implications
of our results, in terms of cluster number counts, are presented in
section \ref{sec:nc}, while the likelihood analysis is shown in
section \ref{sec:likelihood}. Finally, we draw the main conclusions of
this work in section \ref{sec:conc}.

\section{N-body simulations}\label{sec:sims}

For this paper we have used a subset of the large suite of N-body simulations presented in Paper I. We summarize the main features of these simulations here and refer the reader to \cite{Paco_2013} for further details. 

\begin{table}
\begin{center}
\resizebox{15.5cm}{!}{
\begin{tabular}{|c|c|c|c|c|c|c|c|c|c|c|c|}

\hline
Name & \Mnu & Box  & $\Omega_{\rm m}$ & $\Omega_{\rm b}$ & $\Omega_\Lambda$ & $\Omega_\nu$ & $h$ & $n_s$ & $N_\mathrm{CDM}^{1/3}$ & $N_\nu^{1/3}$ & $\sigma_8$ \\
 & [eV] & [$h^{-1}\rm{Mpc}$] & & & & & & & & & $(z=0)$\\
\hline
\hline 
H6 & $0.60$ & $1000$ & 0.2708 & 0.050 & 0.7292 & 0.0131 & 0.7 & 1.0 & $512$ & $512$ & $0.675$\\
\hline
H3 & $0.30$ & $1000$ & 0.2708 & 0.050 & 0.7292 & 0.0066 & 0.7 & 1.0 & $512$ & $512$ & $0.752$\\
\hline
H0 & $0.00$ & $1000$ & 0.2708 & 0.050 & 0.7292 & 0 & 0.7 & 1.0 & $512$ & $0$ & $0.832$\\
\hline
H6s8 & $0.06$ & $1000$ & 0.2708 & 0.050 & 0.7292 & 0.0131 & 0.7 & 1.0 & $512$ & $512$ & $0.832$\\
\hline
\end{tabular}
}
\end{center} 
\caption{Summary of the simulations used in the present work.}
\label{tab_sims}
\end{table}

The N-body simulations have been run using the TreePM code \texttt{GADGET-3}, which is an improved version of the code \texttt{GADGET-2} \cite{Springel_2005}. The neutrinos have been simulated using the so-called \textit{particle-based} implementation (see \cite{Brandbyge_particles, Brandbyge_grid, Brandbyge_hybrid, vhs10, Simeon_2013} for the different methods used to simulate the cosmic neutrino background). 

The starting redshift of the simulations was set to $z=99$. The initial conditions were generated 
at that redshift by displacing the particles positions from a regular cubic grid, using the Zel'dovich approximation. We incorporate the effects of baryons into the CDM particles by using a transfer function that is a weighted average of the transfer functions of the CDM and the baryons, obtained directly from the \texttt{CAMB} code \cite{CAMB}. The Plummer equivalent gravitational softening of each particle type is set to $1/30$ of their mean inter-particle linear spacing. For each simulation we saved snapshots at redshifts 0, 0.5, 1 and 2.

The different cosmological models used for this paper are shown in Table \ref{tab_sims}, together with the values of their cosmological parameters. Each simulation consists of eight independent realizations obtained by generating the initial conditions using different random
seeds. The size of the cosmological boxes are 1 $h^{-1}$Gpc for all the simulations. The 
cosmological models span from a massless neutrino model (H0) to cosmologies with \Mnu = 0.3 eV (H3) and \Mnu = 0.6 eV (H6 and H6s8). The simulations H6, H3 and H0 share
the value of the large--scale power spectrum normalisation $A_s$, whereas the value of this parameter has been tuned in the simulation H6s8 to obtain the same value of $\sigma_8$ of the simulation H0. The values of the other cosmological parameters are common to all the simulations: $\Omega_{\rm m}=\Omega_{\rm cdm}+\Omega_{\rm b}+\Omega_\nu=0.2708$, $\Omega_{\rm b}=0.05$, 
$\Omega_\Lambda=0.7292$, $h=0.7$ and $n_s=1.0$. In all the simulations the value of the
parameter $\Omega_{\rm cdm}$ is given by $\Omega_{\rm m}-\Omega_{\rm b} - \Omega_\nu$, i.e. is fixed by 
requiring that the total matter density of the Universe is the same for all the cosmological models. The number of CDM particles is $512^3$, and for the models with massive neutrinos the number of neutrinos is also $512^3$. The masses of the CDM particles are $5.6\times10^{11}~h^{-1}$ M$_\odot$ for the model H0, while for the others model the masses are slightly different since the value of
$\Omega_{\rm cdm}$ varies from model to model.

\section{The halo mass function} 
\label{sec:massf}

We start this section by explaining how we identify the dark matter haloes from the snapshots of the 
N-body simulations. We then investigate whether the Tinker fitting formula \cite{Tinker2008}
along with the so-called \textit{cold dark matter prescription} reproduces the HMF of N-body
simulations for cosmological models with massive neutrinos.

The dark matter haloes have been identified using the \texttt{SUBFIND} algorithm \cite{Subfind}.
Even though \texttt{SUBFIND} is capable of identifying all the haloes and sub-haloes from a given particle distribution, we have used it to identify spherical overdensity (SO) haloes, which correspond to the groups identified by \texttt{SUBFIND}. The virial radius of a given dark matter halo corresponds to the radius within which the mean density is $\Delta=$200 times the mean density of the Universe. We restrict our analysis to SO haloes containing at least 32 particles.

\texttt{SUBFIND} has only been run on top of the CDM particle distribution. This is equivalent to neglect the contribution of neutrinos to the masses of the dark matter haloes. Such assumption is supported by different studies \cite{Wong, Brandbyge_halos, Paco_Jordi, Paco_2013a} which have shown that the contribution of massive neutrinos to the total mass of dark matter haloes is below the percent level for the neutrino mass range relevant for this paper. We have explicitly checked that the contribution of neutrinos with \Mnu = 0.6 eV to the total masses of dark matter haloes ranges from $0.01\%$ for haloes with $M_{200}\simeq 10^{13}$ $h^{-1}$M$_\odot$ to $0.5\%$ for the most massive haloes with $M_{200}\simeq 10^{15}$ $h^{-1}$M$_\odot$. To make sure that our results are not affected by 
selecting the haloes on top of the CDM particle distribution we have run \texttt{SUBFIND} on top of the total
matter (i.e. CDM plus neutrinos) density field. We find that the HMF of SO haloes changes by less than
$0.5\%$ on a very wide range of masses. However, the masses of some low mass haloes are slightly
changed when including neutrinos. This is because some of these low mass haloes contain many unbound neutrino particles, which bias the estimate of their masses by an unreasonable amount. 
This effect is less important for more massive haloes and/or for simulations in which the number of 
neutrino particles is much larger than the number of CDM particles. In order to avoid this spurious
contamination in the masses of some dark matter haloes we decided to rely on the halo catalogues
obtained by running \texttt{SUBFIND} just on top of the CDM particle distribution.

Now we turn our attention to the halo mass function for cosmologies with massless and massive neutrinos. It is a common practice to parametrize the abundance of dark matter haloes in the following way:
\begin{equation}
n(M,z)=f(\sigma,z)\frac{\rho}{M}\frac{d\log \sigma^{-1}(M,z)}{dM}~;
\label{MF}
\end{equation}
where $n(M,z)$ is the comoving number density of dark matter haloes per unit mass at redshift $z$, 
$\rho$ is the comoving mean density of the Universe and $\sigma(M,z)$ is defined as:
\begin{equation}
\sigma^2(M,z)=\frac{1}{2\pi^2}\int_0^\infty k^2P(k,z)W^2(k,R)dk~,
\label{eq:sig2}
\end{equation}
with $P(k,z)$ being the linear matter power spectrum at redshift $z$, while $W(k,R)$ is the Fourier
transform of the top-hat window function of radius $R$. The relationship between
the halo mass, $M$, and the radius in the top-hat window function is given by $M=4\pi\rho R^3/3$.

Our aim here is to compare the results of the left and right-hand side of eq.~(\ref{MF}). The left-hand side can be directly measured 
from the N-body simulations, whereas the right-hand side can be computed using a fitting formula for the function $f(\sigma,z)$ together with some prescriptions for cosmological models with massive neutrinos. We calculate the left-hand side of eq.~(\ref{MF}) by approximating the quantity $dn(M,z)/dM$ by $\bigtriangleup n(M,z)/\bigtriangleup M$, where the width of the mass intervals has been chosen to be $\bigtriangleup \log(M)=0.2$. The comoving number density of dark matter haloes in a given mass interval $\bigtriangleup n(M,z)$ has been directly obtained from the N-body halo catalogue. In order to compute the right-hand side of eq.~(\ref{MF}) we need the following 
three ingredients: 1) the function $f(\sigma,z)$; 2) the value of $\rho$ to establish the relation between the halo mass and the radius in the top-hat window function and 3) the linear matter power spectrum $P(k,z)$. 

Since we are considering SO haloes, we compare our N-body results to the fitting formula of Tinker et al. \cite{Tinker2008}, also defined in terms of SO haloes. The Tinker fit assumes the functional form proposed by \cite{Warren:2005ey} with 
\begin{equation}
 \label{eq:Tinker}
f(\sigma)=A\left[\left(\frac{\sigma}{b}\right)^{-a}+1\right]e^{-c/\sigma^2}~,
\end{equation}
where $A$, $a$, $b$ and $c$ are the best-fit parameters\footnote{Notice that the Tinker best fit parameters have an explicit redshift dependence, i.e. the Tinker halo mass function is not universal in redshift.} for the overdensity $\Delta=200$ presented in~\cite{Tinker2008}.

In a standard $\Lambda$CDM cosmology the quantities $\rho$ and $P(k)$ appearing in eqs.~(\ref{MF}, \ref{eq:sig2}) are evaluated for the total dark matter field. However, it is not obvious which quantities have to be used for a model with massive neutrinos. The work of \cite{Brandbyge_halos} demonstrated that the abundance of dark matter haloes in massive neutrino cosmologies cannot be reproduced by the ST fit if the {\em total} matter density and linear power spectrum were used when calculating the r.h.s. of eq.~(\ref{MF}). The authors proposed to use, instead, the mean cold dark matter density $\rho_{\rm cdm}$, computing, however, the variance $\sigma^2(M,z)$ still in terms of the total matter power spectrum. Such prescription, that we will refer to as the \textit{matter prescription}, was later corroborated by several works \cite{Marulli_2011, Paco_2013a}.

\begin{figure}
\begin{center}
\includegraphics[width=1.0\textwidth]{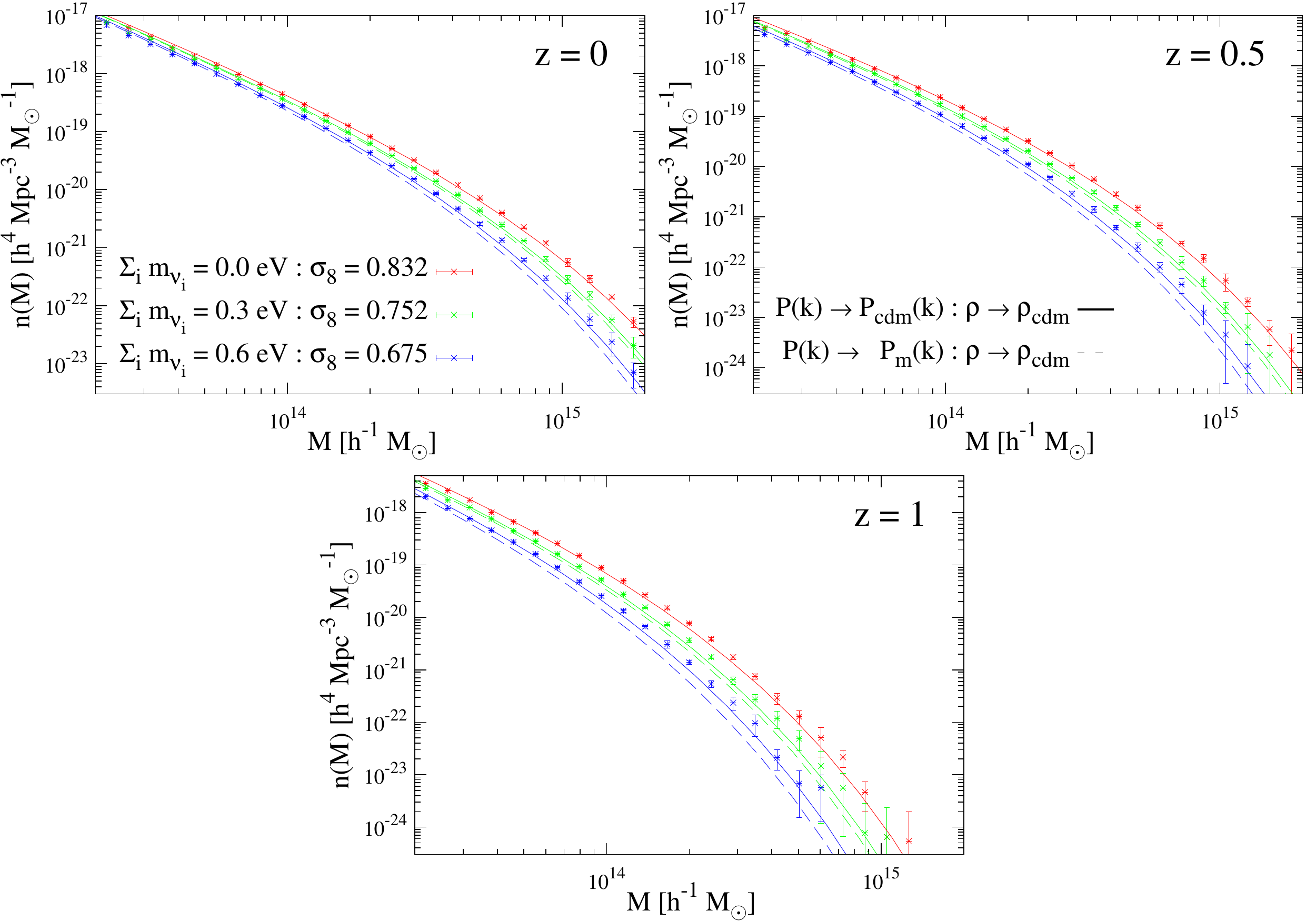}
\end{center}
\caption{Mass function of dark matter haloes identified using the SO criteria
for different cosmological models at redshifts $z=0$ (upper-left), $z=0.5$ (upper-right) 
and $z=1$ (bottom). The points show the halo mass function obtained from the N-body simulations with massless neutrinos (red) and with neutrinos with masses \Mnu = 0.3 eV (green) and \Mnu = 0.6 eV (blue). The error bars represent the dispersion around the mean value obtained from the eight independent realizations for each cosmological model. The results of using the Tinker fitting formula \cite{Tinker2008} along with the matter and cold dark matter prescriptions (see text for details) are  displayed with dashed and solid lines, respectively.}
\label{HMF}
\end{figure}

More recently, the authors of \cite{Ichiki-Takada} studied the gravitational collapse of a spherical region in a massive neutrino cosmology, showing that neutrinos play a negligible role in the process and leading to the conclusion that the cold dark matter power spectrum should be used to predict the halo mass function. Indeed, in Paper II we show that, for Friends-of-Friends (FoF) haloes, a good agreement between the MICE fitting formula \cite{Crocce_2010} and our N-body simulations is obtained if both $\rho$ and $P(k)$ are computed in terms of CDM quantities alone. We call this the \textit{cold dark matter prescription} for massive neutrino cosmologies, and in Paper II we show that it is the only way of obtaining a mass function that is nearly universal with respect to changes in the background cosmology.

We now compare the abundance of dark matter haloes from the N-body simulations with the Tinker prediction evaluated with both the matter and cold dark matter prescriptions. We emphasize that for cosmologies with massless neutrinos the above two prescriptions become the same. We show the results of this comparison in Fig.~\ref{HMF} where the data points correspond to the mean of the mass function, $n(M)$, measured from eight realizations while the error bars represent the error on the mean. Predictions using the Tinker fitting formula along with the matter and cold dark matter prescriptions are shown by the dashed and solid lines, respectively. 
We show the results at redshifts 0, 0.5 and 1 for the simulations H0, H3 and H6  (results at $z>1$ 
are noisy). For clarity we do not display the results of the simulation 
H6s8 since they are very close to those of the simulation H0. 

We find that the cold dark matter prescription reproduces much better the abundance of dark matter 
haloes extracted from the N-body simulations. The agreement between the Tinker fitting formula (plus 
the \textit{cold dark matter prescription} for massive neutrinos) and our results is pretty good at $z=0$, 
while at higher redshift is a bit poorer. We note that the differences between the results from the N-body
 simulations and the Tinker fitting formula along with the cold dark matter prescription are almost
 independent of the cosmological model, likely arising from the different method used to identify the
 SO halos with respect to Tinker et al.~\cite{Tinker2008}. In addition, Paper II shows that the halo mass function for FoF 
haloes ($b=0.2$) in our N-body simulations is very well reproduced (within a $10\%$) by the fitting 
formula of Crocce et al. \cite{Crocce_2010} at all redshifts. 
We emphasize that the use of a different halo mass function will not change the main conclusions of 
this paper.

\begin{figure}
\begin{center}
\includegraphics[width=1.0\textwidth]{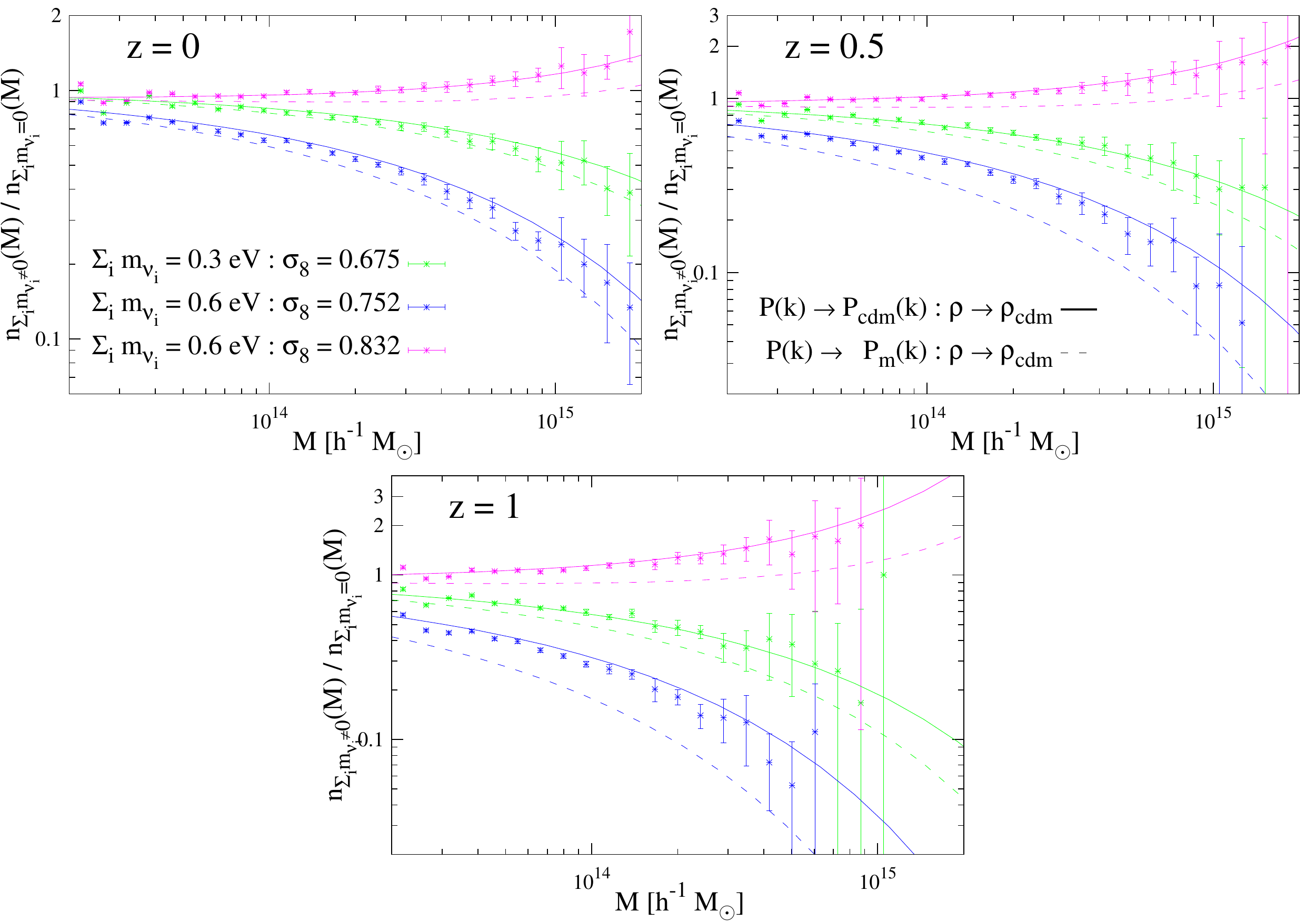}
\end{center}
\caption{Ratio between the halo mass function for cosmologies with massive neutrinos to the halo
mass function for the cosmological model with massless neutrinos (reference model).
The points represent the results
from the N-body simulations whereas the solid and dashed lines correspond to the ratios between
the halo mass functions computed using the Tinker fitting formula together with the \textit{cold dark
matter prescription} and the \textit{matter prescription} for cosmologies with massive neutrinos, 
respectively.}
\label{ratio_HMF}
\end{figure}
 
In Fig.~\ref{ratio_HMF} we show the ratio of the halo mass function for cosmologies with massive
neutrinos to the halo mass function for the cosmology with massless neutrinos. We find that the 
abundance of SO haloes is very well reproduced by the Tinker fitting formula once the cold dark matter prescription is used for cosmologies with massive neutrinos.

\section{An application to cluster number counts}
\label{sec:nc}

A different prescription for the HMF can affect the constraints
on cosmological parameters provided by cluster number counts by changing the number
of clusters predicted for a given cosmology and survey.

\begin{figure}
\centering
\includegraphics[width=0.49\textwidth]{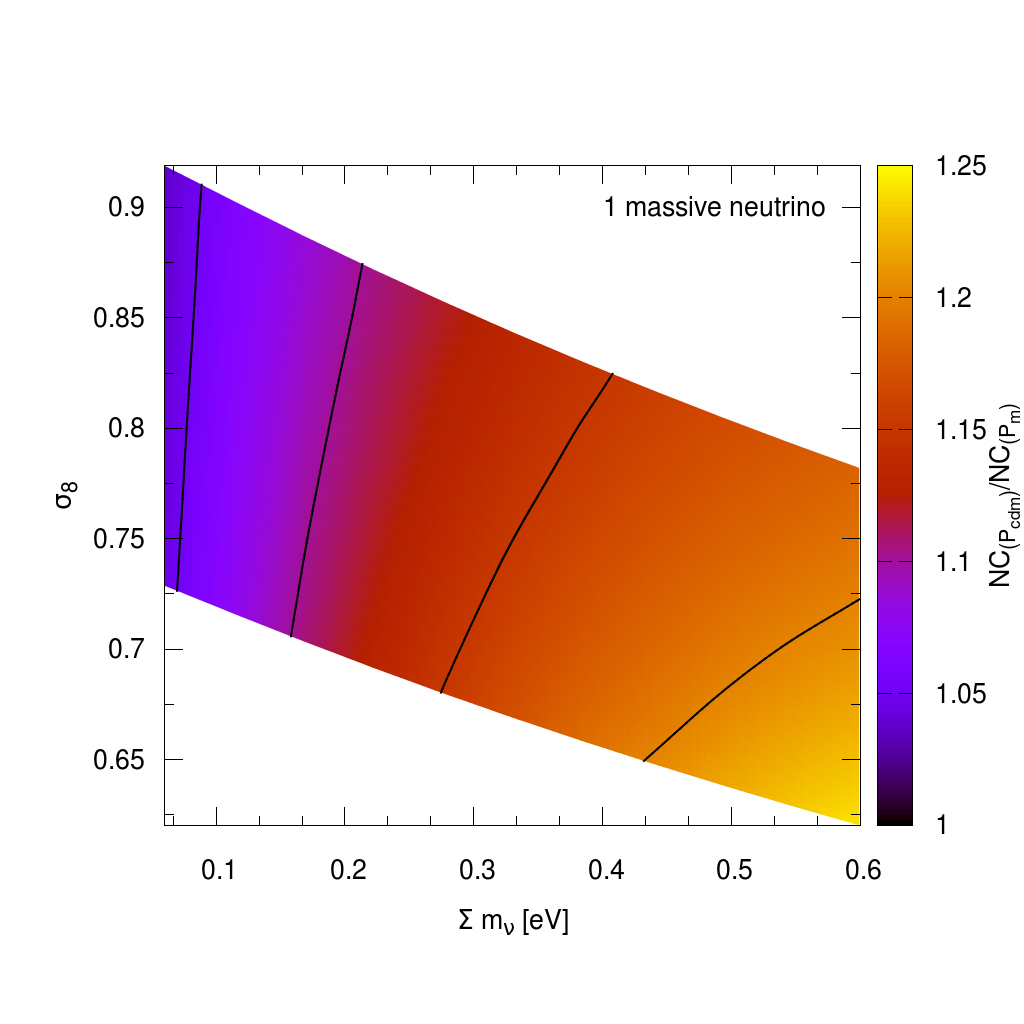}
\includegraphics[width=0.49\textwidth]{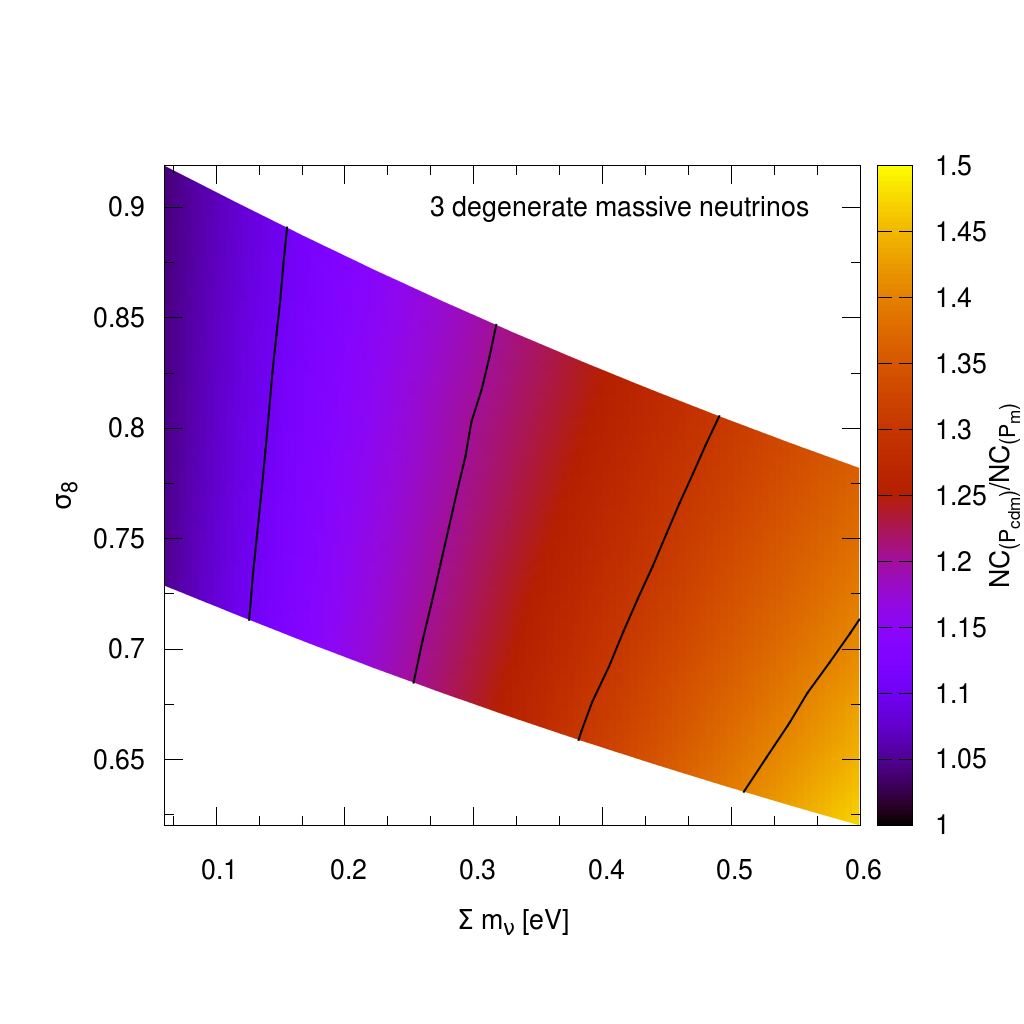}
\caption{Ratio of the number counts obtained using the CDM over the matter prescription for different combinations
 of ($\sum m_\nu - \sigma_8$) values (\textit{colour-coded}) and different neutrino mass splitting: one massive neutrino 
(\textit{left} panel) and three degenerate massive neutrinos (\textit{right} panel).
 For a given cosmology, the cold dark matter prescription predicts a larger number
 of clusters, especially for high neutrino mass and cosmology with three massive neutrinos. Black curves trace constant values of the ratio;
from the left to the right: $1.05, 1.10,1.15,1.20$ (\textit{left} panel); $1.1, 1.2,1.3,1.4$ (\textit{right} panel).}
\label{fig:ncratio}
\end{figure}

The number of cluster expected to be detected within a survey with sky coverage 
$\Delta \Omega$ in a redshift bin $[z_i,z_{i+1}]$ can be expressed as:
\begin{equation}
 N_i = \int_{z_i}^{z_{i+1}} {\rm d}z \int_{\Delta \Omega} {\rm d}\Omega \frac{{\rm d}V}{{\rm d}z{\rm d}\Omega} \int_0^\infty dM \,X(M,z,{\bm \Omega})\,n(M,z)\, ,
\label{eq:nc_z}
\end{equation}
where ${\rm d}V/{\rm d}z{\rm d}\Omega$ is the comoving volume element per unit redshift
and solid angle, $X(M,z,{\bm \Omega})$ is the survey completeness and $n(M,z)$ is the halo mass function. In what
 follows we adopt the Tinker functional form for the mass function defined in eq.~(\ref{eq:Tinker}) 
with the best-fit parameters for the overdensity $\Delta=200$ as provided by~\cite{Tinker2008}.

The completeness function depends on the strategy and specifics of the survey. For the purpose of this work we can simply express this function as
\begin{equation}
 X(M,z,{\bm \Omega})= \int_{M_{\rm lim}(z)}^{\infty}{\rm d}M^{\rm ob} p(M^{\rm ob}\|M)\, ,
 \label{eq:completeness}
\end{equation}
where the lower limit in the mass integral, $M_{\rm lim}(z)$, 
represents the minimum value of the observed mass for a cluster to be included in the survey,
and it is determined by the survey selection function and the fiducial signal-to-noise level adopted. 

The function $p(M^{ob}\|M)$ gives the probability that a
cluster of true mass $M$ has a measured mass given by $M^{ob}$ and
takes into account the uncertainties that a scaling relation
introduces in the knowledge of the cluster mass. Under the
assumption of a lognormal-distributed intrinsic scatter around the
nominal scaling relation with variance $\sigma^2_{\ln M}$, the
probability of assigning to a cluster of true mass $M$ an observed mass
$M^{ob}$ can be written as~\cite{LimaHu2005}:
\begin{equation}
p(M^{\rm ob}\|M)\,=\,\frac{1}{M^{\rm ob}  \sqrt{\left( 2\pi \sigma^2_{\ln M}\right) }} \exp \left[-\frac{(\ln M^{\rm ob}-B_{\rm M}-\ln M)^2}{2 \sigma^2_{\ln M}}\right] \,,
\label{eq:prob}
\end{equation}
where the parameter $B_{\rm M}$ represents the fractional value of the
systematic bias in the mass estimate.  

We now turn to the implications of the prescription choice on the HMF prediction.
By replacing $P_{\rm m}(k,z)$ with $P_{\rm cdm}(k,z)$ one neglects the suppression of 
the total DM density fluctuations on scales smaller than their free-streaming length,
the scale below which neutrinos cannot cluster due to their high thermal velocity
(see, e.g.~\cite{2006PhR...429..307L}). The magnitude of the suppression depends 
on the sum of the neutrino masses, while the scale below which it takes place
depends on the individual neutrino mass and on redshift.  
This in turn affects the halo mass function by shifting the maximum cluster mass
(i.e. the scale beyond which the halo mass function is exponentially suppressed)
to larger values, thus increasing the predicted number of rare massive clusters.
The effect is larger for larger total neutrino mass, larger number
of massive neutrinos and higher redshift. 

In Fig.~\ref{fig:ncratio} we show the ratio of the cluster counts predicted
using the $P_{\rm cdm}(k,z)$ prescription over the one predicted using 
$P_{\rm m}(k,z)$ (colour coded) for different combinations of ($\sum m_\nu - \sigma_8$)
values and for two neutrino mass split schemes: a single massive neutrino (left panel)
and three degenerate massive neutrinos (right panel). 
In the former case the total neutrino mass is assigned to one neutrino species ($m_1= \sum m_\nu$ and $m_2=m_3=0$), 
in the latter one it is equally split among three massive species ($m_{\rm i}= \sum m_\nu /3$ with $i=1,2,3$).
The plots have been obtained by varying $\sum m_\nu$ and $A_s$ and keeping fixed
$\Omega_{\rm m}$, $\Omega_{\rm b}$, $\tau$, $H_0$ and $n_s$ to the Planck mean value 
(\cite{Planck_cosm_parameters}; Table 2, \textit{Planck}+WP).
In order to mimic a Planck SZ-cluster survey,
we computed the number counts integrating eq.~(\ref{eq:nc_z}) between
$0.0<z<1.0$ with a sky coverage $\Delta \Omega =27.000~\text{deg}^2$
and we approximated the Planck SZ-cluster completeness function
using as lower limit in eq.~(\ref{eq:completeness}) the limiting mass $M_{\rm lim}(z)$\footnote{Following the 
recipe given in~\cite{HuKravtsov2003}, the limiting mass has been 
converted to $M_{\rm lim,200}(z)$ -- the limiting mass within a radius
encompassing an overdensity equal to 200 times the mean density of 
the Universe -- consistently with the chosen halo mass function.}
provided by the Planck Collaboration (dashed black line in Fig.~3 of \cite{PlanckSZ2013}).
Moreover, since we are simply interested in quantify the relative effect of
 using an improved HMF calibration we assumed no uncertainties 
in the estimation of the true mass ($M=M^{\rm ob}$) and we set
 $B_{\rm M}=0$ and $\sigma^2_{\ln M}\rightarrow0$ in eq.~(\ref{eq:prob}). 
Power spectra have been computed using {\tt CAMB}~\cite{CAMB}, where
$P_{\rm cdm}(k,z)$ has been obtained exploiting the relation
\begin{equation}
 P_{\rm cdm}(k,z)=P_{\rm m}(k,z) \left(\frac{\Omega_{\rm cdm} T_{\rm cdm}(k,z)+\Omega_{\rm b} T_{\rm b}(k,z)}{\Omega_{\rm cdm}+\Omega_{\rm b}} \frac{1}{T_{\rm m}(k,z)}\right)^2\, ,
\label{eq:p_cdm}
\end{equation}
with $T_{\rm cdm}$, $T_{\rm b}$ and $T_{\rm m}$ being the CDM, baryon and total matter transfer functions, respectively.

Assuming one massive neutrino, changing the matter power spectrum
to the cold dark matter one in the HMF prediction increases the
expected number of clusters by $\sim5\%$ in the minimal normal hierarchy scenario ($\sum m_\nu=0.06$eV), 
reaching differences of $\sim 20\%$ for masses of $\sum m_\nu \sim 0.4 {\rm eV}$.
Considering instead three degenerate massive neutrinos,
the CDM prescription gives even a larger correction to the cluster counts:
the splitting of the total neutrino mass between three species
causes the free-streaming length to increase, therefore widening
the  range in which $P_{\rm m}(k,z)$ is suppressed with respect
to $P_{\rm cdm}(k,z)$. As a result, the difference in
cluster counts computed with the two prescriptions
reaches $\sim30\%$ for neutrino mass of the order of $\sum m_\nu = 0.4 {\rm eV}$.  
For a given cosmology the magnitude of the ratio slightly
depends also on the specifics of the survey: a lower $M_{\rm lim}(z)$
would entail a larger difference between the expected number of 
clusters computed with the two different calibrations.

\begin{figure}
\includegraphics[width=7.3cm]{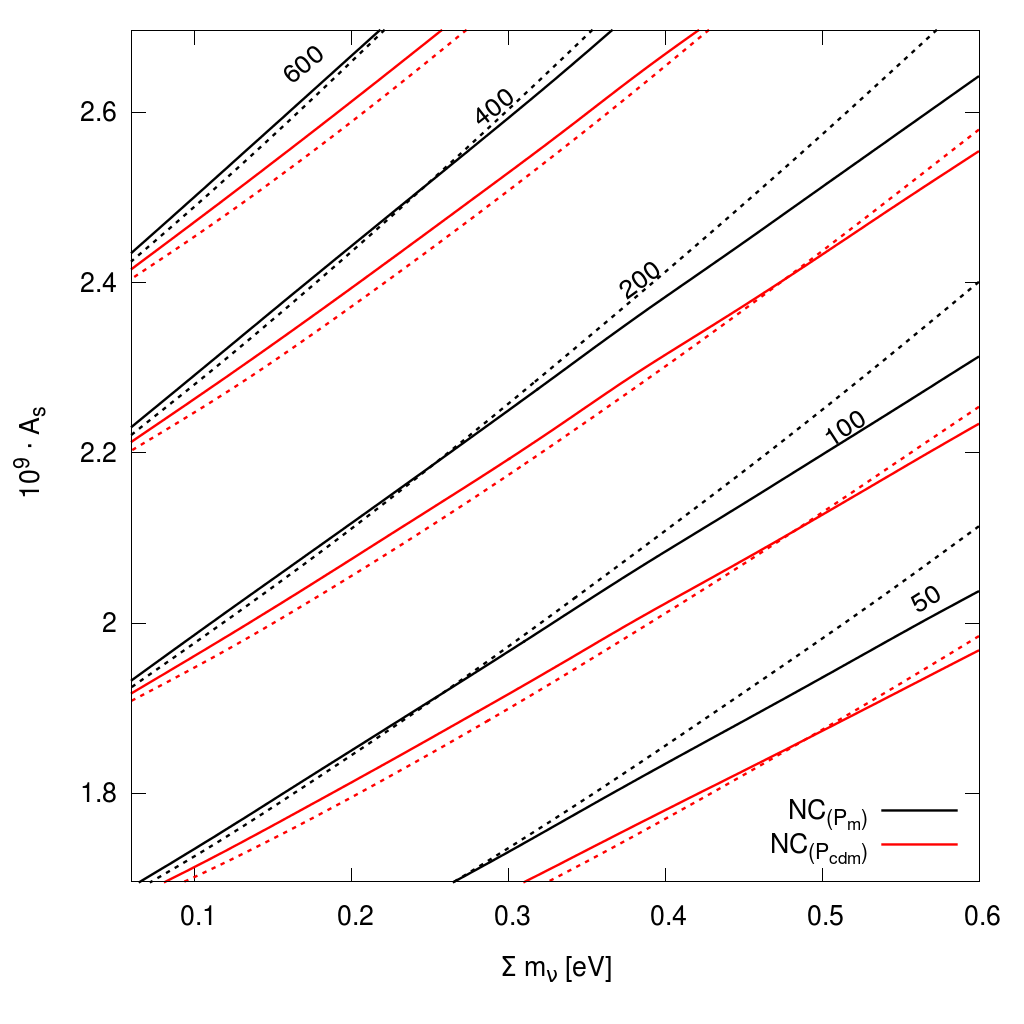}
\includegraphics[width=7.3cm]{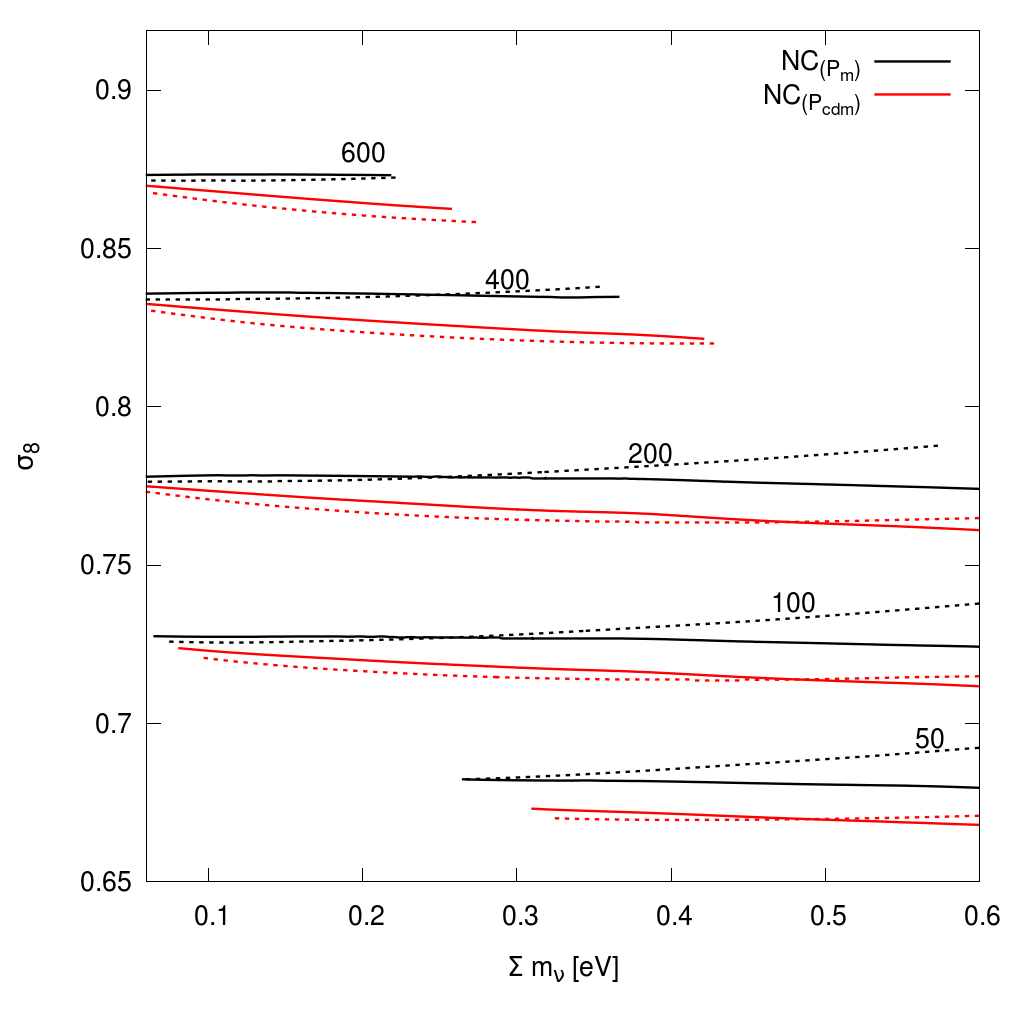}
\caption{Curves of constant number counts ($N=600$, 200, 100 and 50, top to bottom)
in the plane $\sum m_\nu - 10^9 \cdot A_s$ (\textit{left panel}) and in the plane
$\sum m_\nu$ - $\sigma_8$ (\textit{right panel}), for the two prescriptions for
the halo mass function, matter (\textit{black}) and cold dark matter (\textit{red}) and for
two neutrino mass splitting schemes, single massive neutrino (\textit{solid lines})
and three degenerate massive neutrinos (\textit{dashed lines}). The different slope
of the black and red curves shows the different degeneracy direction between parameters in the prescriptions.}
\label{fig:nciso}
\end{figure}

The difference in the predictions in turn affects 
the degeneracy between cosmological parameters.
An example of this effect is shown in figure~\ref{fig:nciso},
in the ($\sum m_\nu - A_s$) plane (left panel) and the
corresponding ($\sum m_\nu-\sigma_8$) plane (right panel). The curves
correspond to constant number counts 
obtained using $P_{\rm m}(k,z)$ (black) or $P_{\rm cdm}(k,z)$ (red)
in the halo mass function definition, following
the same procedure of figure~\ref{fig:ncratio} to compute the expected number of clusters
and keeping the other cosmological parameters ($\Omega_{\rm m}$, $\Omega_{\rm b}$, $\tau$, $H_0$, $n_s$) fixed to the Planck mean value.
Solid and dashed curves are for models with one massive neutrino and
three degenerate massive neutrinos, respectively. The different slope
of the curves indicates a different degeneracy direction between parameters.
Consistently with the results shown in figure~\ref{fig:ncratio} the change in the slope
is more pronounced in the case of three massive neutrinos. 

As illustrated in the next section both these effects can contribute to modify
the information on cosmological parameters inferred from cluster data in models
with massive neutrinos.

\section{Implications for cosmological constraints}
\label{sec:likelihood}

The ultimate aim of an analytic expression for the halo mass function
is to link the observed abundance of galaxy clusters to the underlying cosmology.
The recently released Planck data indicate some tension between
the cosmological parameters preferred by the primary CMB temperature measurements and
those obtained from cluster number counts using X-ray~\cite{2009ApJ...692.1060V},
optical richness~\cite{2010ApJ...708..645R} and SZ-selected clusters~\cite{SPT2013,ACT2013,PlanckSZ2013}.
In particular, the values of $\sigma_8$ and $\Omega_{\rm m}$ inferred
from cluster analyses are found to be lower than the values derived from CMB data.
It has been argued that this discrepancy could be due to a wrong calibration of 
cluster mass (see e.g.~\cite{Henry2009,Rozo2013}) or alternatively it may indicate the need to extend the
minimal $\Lambda$CDM to massive neutrinos~\cite{PlanckSZ2013,Wyman2013,Battye2013,Hamann2013}. 
In the latter case, the results presented in this paper could in principle affect
derived cosmological constraints which relies on an incorrect calibration of the HMF
in cosmological models with massive neutrinos.
In fact, in all previous cluster studies, the variance of the total 
dark matter field has been used to put constraints on neutrino masses.
In section~\ref{sec:nc} we have shown that, given a background cosmology, using the \textquotedblleft wrong\textquotedblright
prescription for the HMF could lead to differences up to $30\%$ in the expected number of cluster.
However that is not the the only reason to use the variance of the CDM field.
Indeed, a key assumption in previous cosmological analyses of clusters counts is that the shape of the HMF is independent of the underlying cosmology, and the same functional form can be used through all the parameter space. In Paper II we show that universality of the HMF with respect to neutrino masses, and more in general cosmology, is recovered only if the cold dark matter prescription is adopted. This is another important effect that should be taken into account by future studies.

In order to assess the effects of the cold dark matter prescription on the parameter
estimation we choose as a case study the sample of 188 SZ-selected
clusters with measured redshift published in the Planck SZ Catalogue~\cite{PlanckSZ_catalogue}.
The cosmological constraints have been obtained using the likelihood function for
Poisson statistics~\cite{Cash1979}: 
\begin{equation}
  \ln L(N^{\rm obs} | N^{\rm th}) = \sum_{i=1}^{N_{\rm bin}} \left[ N_i^{\rm obs} \ln(N_i^{\rm th}) - N_i^{\rm th} - \ln(N_i^{\rm obs}!) \right]\,,
  \label{eq:like}
\end{equation}
where $N_i^{\rm obs}$, $N_i^{\rm th}$ represent respectively
the number of clusters observed and theoretically predicted
in the $i$-th redshift bin. The redshift range has been 
divided in $N_{\rm bin}=10$ bins of width $\Delta z=0.1$
between $z=0.0$ and $z=1.0$, also including in the
analysis redshift bins with no observed clusters. 
We computed the predicted number of clusters , $N_i^{\rm th}$, for a Planck-like SZ-cluster survey
following the procedure described on section~\ref{sec:nc}. The parameter
space has been explored by means of the Monte Carlo Markov Chain technique using the publicly available code
{\tt CosmoMC}\footnote{http://cosmologist.info/cosmomc/}~\citep{Lewis2002},
where we included a module for the computation of the likelihoods function described above.
Since we are interested in the effects that a different prescriptions
for the HMF has on parameter constraints rather than the constraints themselves,
we kept fixed $\Omega_{\rm b}h^2=0.022$, $\tau=0.085$ and $n_s=0.963$,
allowing only $\Omega_{\rm cdm}h^2$, $\theta$, $\log(10^{10} \cdot A_s)$ and $\sum m_\nu$ to vary.
\begin{figure}
\centering
\includegraphics[width=1.0\textwidth]{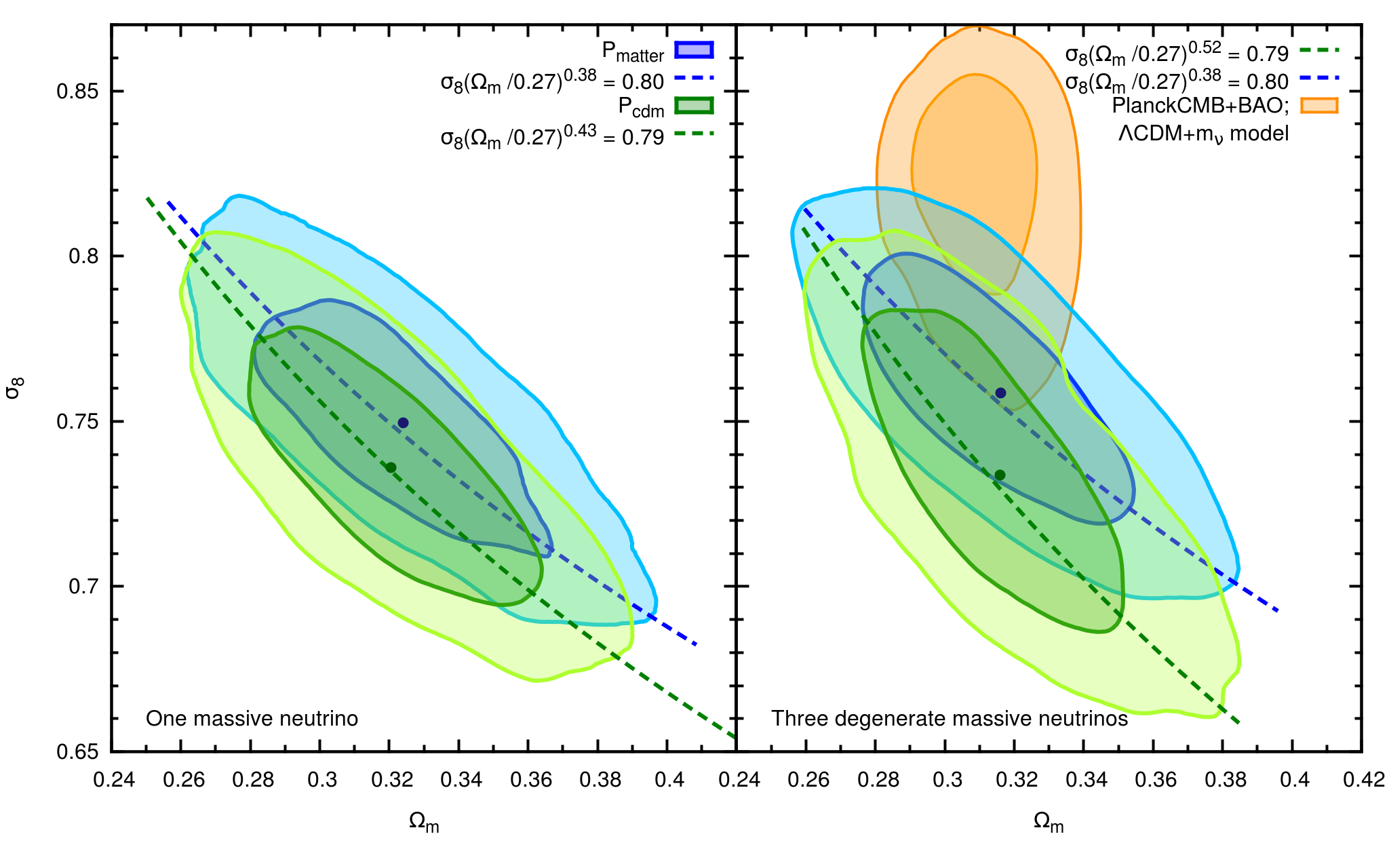}
\caption{Comparison of the $68\%$ and $95\%$ C.L. contours in the 
$\Omega_{\rm m}-\sigma_8$ plane and degeneracy curves obtained using the matter (\textit{blue})
and cold dark matter (\textit{green}) prescription. We show the results when the sum of the neutrino
masses is split between one massive neutrino family (\textit{left panel}) and three degenerate
neutrino families (\textit{right panel}). Also shown in the
\textit{right} panel in \textit{orange} the contours from 
PlanckCMB+BAO datasets for a $\Lambda$CDM+$\sum m_\nu$ model.}
\label{fig:oms8}
\end{figure}
For the same reason we neglect errors on nuisance parameters, 
which have been kept fixed to $B_{\rm M}=0$ and $\sigma^2_{\ln M}=0.2$ in order
to roughly reproduce the mean values of $\Omega_{\rm m}$ and $\sigma_8$ obtained
by the Planck Collaboration with Planck-SZ+BAO+BBN data~\cite{PlanckSZ2013}.
We also checked that our results in the $\Omega_{\rm m}-\sigma_8$ plane fixing $\sum m_\nu$ to $0.06 {\rm eV}$
were in good agreement with those obtained by the Planck Collaboration. 
Finally, due to the weak sensitivity of the clusters sample to some cosmological parameters, we set 
a Gaussian prior on the total neutrino mass, $\sum m_\nu = 0.06 \pm 1.0 {\rm eV}$, and one on the expansion
rate, $H_0 = 68.9 \pm 3.0$ km/s/Mpc (from BAO measurements~\cite{Addison2013}).
Note that the actual Planck cluster likelihood is not publicly
available. Therefore, a quantitative comparison
with the SZ Planck cluster  results is not possible. However, since
we are presenting results in terms of relative effects between different HMF calibrations,
we expect that our findings will be robust and could be quantitatively similar to those
to be derived with a more accurate likelihood analysis.

The joint constraints on the $\Omega_{\rm m}$-$\sigma_8$ plane resulting
from this analysis are shown in figure~\ref{fig:oms8} with green contours
for the CDM prescription and blue contours for the matter prescription.
The left panel is for a model with one massive neutrino while
the right one for a model with three degenerate massive neutrinos.
The dashed lines show the $\sigma_8 (\Omega_{\rm m}/027)^{\gamma} = S_8$ relation with 
$\gamma$ and $S_8$ parameters obtained by fitting a power-law relation to the likelihood contours.
Also shown in the right panel with orange contours are the constraints from Planck+WP+BAO datasets
for a $\Lambda$CDM cosmology with free $\sum m_\nu$~\footnote{Chains publicly available at http://www.sciops.esa.int/.}. 
While the constraining power of clusters is almost unaffected by different HMF prescriptions the
degeneracy direction become steeper in the CDM case. For one massive neutrino the shift can be quantified
as $\Delta \gamma = 0.05$, or $\Delta \sigma_8 = 0.01$. The effect is even
larger when considering three massive neutrino, for which we obtain a
shift of $\Delta \gamma = 0.14$  and $\Delta \sigma_8 = 0.02$. 
The different degeneracy of the CDM-case contours can be understood as follows:
for our set of free parameters moving toward large $\Omega_{\rm m}$ values in order to keep constant the number of clusters
one has to compensate with lower $\sigma_8$ and larger $\sum m_\nu$ values.
Using the CDM prescription, however, for a given matter density and neutrino mass
the value of $\sigma_8$ which reproduces the right number of cluster is smaller than 
the one recovered using the matter prescription; moreover, the difference between $\sigma_8$
values inferred from the two HMF prescriptions increases with the total neutrino mass and it is more pronounced 
when assuming three massive neutrinos (see Fig.~\ref{fig:nciso}).

Using the orange contours as a reference one can see that the
shift of the contours caused by the CDM prescription goes 
in the direction of increasing the tension with the Planck+BAO results.  
This means that when using the CDM prescription in
trying to reconcile the Planck CMB measurements with cluster 
number counts, when extending the $\Lambda$CDM model
to massive neutrinos, a larger $\sum m_\nu$ value
will result from the combination of the two datasets.

The effects of the usage of the CDM prescription on parameter
estimation are clearly visible but with low statistical significance
for the cluster sample chosen for this work. 
However, owing to the much stronger constraining power expected from
upcoming and future cluster surveys,
corrections to the $\sigma_8$-$\Omega_{\rm m}$ degeneracy direction of the order of $\Delta \gamma \sim 0.1$
would offsets the resulting constraints by a statistically significant amount~\cite{Costanzi2013,Khedekar2013}.

\section{Summary and perspectives}
\label{sec:conc}

By using a set of large box-size N-body simulations containing CDM 
and neutrinos particles we have studied the abundance of dark matter
haloes, identified using the SO criterion, in cosmological models with 
massive neutrinos. The SO haloes have been extracted from the N-body 
simulations by running the \texttt{SUBFIND} algorithm on top of the CDM 
particle distribution to avoid spurious mass contamination in the low mass haloes
from unbounded neutrino particles. We have however
explicitly checked that our results do not change if \texttt{SUBFIND} is run on top
of the total matter density field. We have compared the abundance
of dark matter haloes in cosmologies with massless and massive 
neutrinos with the Tinker fitting formula along with the \textit{matter prescription} and the
\textit{cold dark matter prescription}. In both prescriptions
we use $\rho_{\rm cdm}=\rho_{\rm m}-\rho_\nu$ instead of $\rho_{\rm m}$ when setting the relation 
between the halo mass and the radius in the top-hat window function: $M=4\pi\rho_{\rm cdm}R^3/3$. 
However, in the \textit{cold dark matter prescription} we use the CDM linear
power spectrum, $P_{\rm cdm}$, when computing the value of $\sigma(M,z)$, whereas in the
\textit{matter prescription} we use the total matter linear power spectrum, $P_{\rm m}$. 

We find that the abundance of SO haloes is much better reproduced by the Tinker fitting formula
once the \textit{cold dark matter prescription} is used, in
agreement with the claims of Ichiki \& Takada \cite{Ichiki-Takada} and the results of Paper II \cite{Castorina_2013}. The agreement is very good at $z=0$ while it
worsens a bit at higher redshift.
Once we present the results as ratios of the
halo mass functions for cosmologies with massive neutrinos to the halo mass function for 
cosmologies with massless neutrinos the agreement with theoretical 
predictions improves significantly at all redshifts. We stress that the conclusions
of this paper are not affected if a different halo mass function fitting formula was used. 

We have investigated the effects that the \textit{cold dark matter prescription}
has on theoretically predictions of number counts and on
the estimation of cosmological parameter from cluster samples.
By using the Tinker fitting formula for the HMF we computed the expected number of
clusters for a Planck-like SZ-cluster survey. We found that for a cosmology
with massive neutrinos the predicted number of clusters is higher when using the
cold dark matter prescription with respect to the results obtained by using the matter
prescription.
For a given value of $\Omega_{\rm m}$ the effect is more pronounced for large neutrino 
masses and in the case of a splitting of the total neutrino mass between three degenerate species.
Assuming one massive neutrino family (and two massless neutrino families)
the difference in the predicted number counts between the
two prescriptions is nearly $20\%$ for $\sum m_\nu \sim 0.4 {\rm eV}$,
while it reaches $\sim30\%$ in models with three degenerate massive neutrinos.

The different prediction for the HMF in turn affects the degeneracy
direction between cosmological parameters and the mean 
values inferred from the cluster sample. To quantify these
effects we use as a case study the Planck sample of 188 SZ-selected clusters with measured redshifts.
We performed a Monte Carlo Markov Chains analysis for the parameters 
$\Omega_{\rm cdm}h^2$, $\theta$, $\log(10^{10} \cdot A_s)$ and $\sum m_\nu$,
both splitting the sum of the neutrino masses between one and three massive species.
Looking at the combination $\sigma_8(\Omega_{\rm m}/0.27)^\gamma$, the cold dark matter
prescription provides a steeper degeneracy direction (higher $\gamma$) which causes the
$\sigma_8$ mean value to lower. The shift can be quantified as $\Delta \gamma = 0.05$
and $\Delta \gamma = 0.14$ for one and three massive neutrino respectively,
or in terms of the $\sigma_8$ mean value as $\Delta \sigma_8=0.01$ and $\Delta \sigma_8=0.02$.
The offset has a low statistical significance for the cluster sample used in this work
but it could entail a significant correction when the sample is
combined with other probes or for large cluster samples that
will be provided by future cluster surveys. 
Furthermore, taking into account such an effect has the consequences of exacerbating the tension
between the cosmological parameters derived from CMB data and those of cluster number counts~\cite{PlanckSZ2013}.

\section*{Acknowledgments} 

MC thanks Barbara Sartoris for the useful discussions.
FVN thanks Weiguang Cui for helpful discussions about the
identification of dark matter haloes. The calculations for this paper
were performed on SOM2 and SOM3 at IFIC and on the COSMOS Consortium
supercomputer within the DiRAC Facility jointly funded by STFC, the
Large Facilities Capital Fund of BIS and the University of Cambridge,
as well as the Darwin Supercomputer of the University of Cambridge
High Performance Computing Service (http:// www.hpc.cam.ac.uk/),
provided by Dell Inc. using Strategic Research Infrastructure Funding
from the Higher Education Funding Council for England.  This work has
been supported by the PRIN-INAF09 project ``Towards an Italian Network
for Computational Cosmology'', by the PRIN-MIUR09 ``Tracing the growth
of structures in the Universe'', by the PD51 INFN grant and by
  the Marie Curie Initial Training Network CosmoComp
  (PITN-GA-2009-238356) founded by the European Commission through the
  Framework Programme 7. FVN and MV are supported by the ERC
Starting Grant CosmoIGM. J.-Q. X. is supported by the National Youth
Thousand Talents Program and the Grants No. Y25155E0U1 and
No. Y3291740S3. ES was supported in part by NSF-AST 0908241.


\bibliographystyle{JHEPb}
\bibliography{Bibliography}

\end{document}